\documentclass[12pt]{article}

\usepackage{epsfig}
\usepackage{setspace}
\usepackage{longtable}
\usepackage{amsmath}

\def\e{\begin{equation}}
\def\f{\end{equation}}
\def\=#1{\overline{\overline #1}}

\def\_#1{{\bf #1}}

\def\E{\epsilon}
\def\M{\mu}

\def\.{\cdot}

\def\##1{{\bf#1\mit}}

\def\am{\left(\begin{array}{c}}
\def\amm{\left(\begin{array}{cc}}
\def\a{\end{array}\right)}

\title{Magneto-Dielectric Substrates in Antenna Miniaturization: Potential and Limitations}

\author{Pekka~Ikonen$^{1}$, Konstantin N.~Rozanov$^{2}$, Alexey V.~Osipov$^{2}$, Sergei A.~Tretyakov$^{1}$}

\date{$^{1}$Radio Laboratory/SMARAD, Helsinki University of
Technology\\P.O. Box 3000, FI-02015 TKK, Finland\\
\vspace{0.3cm} $^{2}$Institute for Theoretical and Applied Electromagnetics, 13/19 Izhorskaya ul. Russian
Academy of Sciences, 125412 Moscow, Russia}

\begin{document}

\maketitle {\center \large

Address for correspondence:

Pekka Ikonen \\ Radio Laboratory, Helsinki University of Technology\\ P.O. Box 3000, FI-02015 TKK, Finland

Fax: +358-9-451-2152

E-mail: pekka.ikonen@tkk.fi

}

\parindent 0pt
\parskip 7pt

\vspace{0.5cm}

\begin{center}
\section*{Abstract}
\end{center}

In the present paper we discuss antenna miniaturization using magneto-dielectric substrates. Recent results
found in the literature reveal that advantages over conventional dielectric substrates can only be achieved if
natural magnetic inclusions are embedded into the substrate. This observation is revised and the physical
background is clarified. We present a detailed discussion concerning magnetic materials available in the
microwave regime and containing natural magnetic constituents. The effects of magnetic dispersion and loss are
studied: constraints on the microwave permeability are used to estimate the effect of magnetic substrates on the
achievable impedance bandwidth. Microwave composites filled with thin ferromagnetic films are considered as a
prospective antenna substrate. We calculate the impedance bandwidth for a $\lambda/2$-patch antenna loaded with
the proposed substrate, and challenge the results against those obtained with conventional dielectric
substrates. It is shown that the radiation quality factor is strongly minimized with the proposed substrate even
in the presence of realistic losses. Estimates for the radiation efficiency are given as a function of the
magnetic loss factor.

\textbf{Key words:} Patch antenna, magneto-dielectrics, ferromagnetic film, frequency dispersion, quality
factor, miniaturization

\newpage

\section{Introduction}

It is well known that the impedance bandwidth of electrically small antennas is roughly proportional to the
volume of the antenna in wavelengths. This means that when the antenna is downsized the impedance bandwidth
suffers. However, the trend in the communications industry is towards smaller size devices, thus, efficient
antenna miniaturization is a clear necessity. One of the most common miniaturization techniques is the loading
of the antenna volume with different materials \cite{Bahl}--\cite{Balanis}. Most traditionally, high
permittivity dielectrics have been used to decrease the physical dimensions of the radiator,
e.g.~\cite{Mongia,Hwang,Colburn}. Common problems encountered with high permittivity substrates include e.g.~the
excitation of surface waves leading to lowered radiation efficiency and pattern degradation, and difficulties in
the impedance matching of the antenna. Recently, artificial high-permeability materials working in the microwave
regime \cite{Kostin}--\cite{Maslovski} have been proposed to decrease the size and/or to enhance the impedance
bandwidth properties of microstrip antennas \cite{Hansen}--\cite{Pekka}.

According to the work of Hansen and Burke \cite{Hansen}, inductive (magnetic) loading leads to an efficient size
miniaturization of a microstrip antenna. When the material parameters of the antenna substrate are
\emph{dispersion-free}, and $\mu_{\rm eff}\gg\E_{\rm eff}, \mu_{\rm eff}\gg$~1, a transmission-line (TL) model
for a normal $\lambda/2$-patch antenna predicts that the impedance bandwidth is retained after miniaturization
\cite{Hansen}. Ikonen \emph{et al.}~\cite{Pekka} conducted extensive study on the effect of frequency dispersion
of \emph{artificial}\footnote{A mixture of electrically small, resonating metal unit cells is used to enhance
the magnetic effect.} magneto-dielectric substrates on the impedance bandwidth properties of the loaded antenna.
It was analytically and experimentally shown that due to frequency dispersion of the substrate the obtained
radiation quality factor $Q_{\rm r}$ always exceeds $Q_{\rm r}$ obtained using conventional dielectric
substrates. This conclusion is valid when the voltage and current distribution on the antenna element are not
changed, and the static permeability equals unity. According to the Rozanov limit \cite{Rozanov} for the
thickness to bandwidth ratio of radar absorbers, the thickness of the absorber at \emph{microwave frequencies}
(with a given reflectivity level) is bounded by the \emph{static} value of $\mu$ of the absorber. Discussion
presented in \cite{Pekka} revealed that the static properties of the magneto-dielectric substrate play an
important role also in antenna design. It was shown that a substrate obeying the modified Lorentzian type
dispersion rule and static permeability exceeding unity was advantageous in terms of minimized $Q_{\rm r}$ over
some frequency range.

The present work has two main goals: 1) give a simple physical explanation on the advantages of magnetic
materials in antenna miniaturization by considering the actual measurable quantities on the antenna element,
namely the resonant current and voltage amplitudes. Moreover, using a lumped element model we explicitly reveal
the effect of the static permeability and frequency dispersion. The presently obtained results are generalized
by revising the main results obtained in \cite{Pekka}. 2) Present discussion concerning magnetic materials
available in the microwave regime and containing natural magnetic inclusions. Microwave composites filled with
thin ferromagnetic films \cite{Lagarkov_mod} are introduced as prospective antenna substrates. The effects of
magnetic dispersion and loss are studied. Using a TL-model we calculate the impedance bandwidth properties of a
patch antenna loaded with the proposed substrate and a conventional dielectric substrate. It is shown that in
terms of minimized radiation quality factor the proposed substrate outperforms conventional dielectric
substrates. Estimates for the radiation efficiency are given as a function of the magnetic loss factor.

The rest of the paper is organized in the following way: in section II we revise the effect of the antenna
substrate on $Q_{\rm r}$. Section III presents discussion concerning feasible magnetic materials and discusses
constraints on the achievable microwave permeability. The calculated impedance bandwidth characteristics are
presented in Section IV. The work is concluded in Section V.

\section{The effect of antenna substrate on $Q_{\rm r}$}

\subsection{Physical insight using dispersion-free material parameters}

\begin{figure}[b!]
\centering \epsfig{file=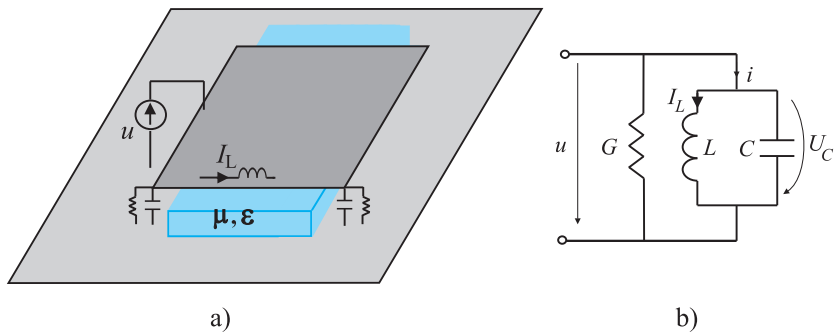, width=11cm} \caption{a) Schematic illustration of a patch antenna fed with
a voltage source. b) Equivalent circuit representation in the vicinity of the fundamental resonant frequency.}
\label{circuit}
\end{figure}

It is known that a magnetic substrate is advantageous in microstrip antenna miniaturization if the material
parameters of the substrate are lossless and \emph{dispersion-free}, e.g.~\cite{Hansen}. To give a physical
explanation on this phenomenon let us consider a resonant $\lambda/2$-patch antenna operating at a fundamental
angular frequency $\omega_{\rm x}$. Moreover, assume the antenna to be filled with hypothetical dispersion-free
and lossless materials characterized by $\mu,\E$, Fig.~\ref{circuit}. The impedance behavior of the antenna can
be approximated with the behavior of a parallel-resonant circuit in the vicinity of $\omega_{\rm x}$. Using the
lumped element representation we get for the quality factor of the antenna \e Q = \frac{1}{G\omega_{\rm
x}L_0\mu}=\frac{C_0\E\omega_{\rm x}}{G}, \label{Qcircuit} \f where $G$ represents the losses in the resonator
(in this case mainly due to radiation), and $L_0$ and $C_0$ are the inductance and the capacitance of the empty
resonator. From (\ref{Qcircuit}) we draw the known observation: with a fixed $\omega_{\rm x}$ an increase in the
inductance (permeability of the filling material) lowers the quality factor, whereas increase in the capacitance
(permittivity) increases the quality factor leading to a narrow impedance bandwidth.

A heuristic idea of the physics behind this observation can be received by considering the amplitude of the
current $|I_L|$ oscillating in the circuit. At the resonance this amplitude is typically very strong. When
thinking of the antenna structure presented in Fig.~\ref{circuit}a.~the strong current flowing in the antenna
element (and in the ground plane) creates a flux which only contributes to stored energy, since radiation
happens at the edges of the patch. Moreover, if one leaves the ends of the patch free from the filling (as is
done in Fig.~\ref{circuit}a.) the filling has only a very small effect on the radiation conductance. So
intuitively one should try to decrease the amplitude of the oscillating current as much as possible. Increase in
$C$ will lower the impedance of the capacitor, moreover at the same time we must decrease $L$ (to keep
$\omega_{\rm x}$ fixed) and this further leads to decreased impedance level. Thus, it will be easy for the
oscillation to build up in the circuit. The same observation will be seen by considering the stored energy in
the circuit at the resonance \e W = \ W_{\rm e} + W_{\rm m} = \frac{L}{2}|I_L|^2 + \frac{C}{2}|U_C|^2 =
\frac{|U_C|^2}{2}\bigg{(}\frac{1}{\omega^2L}+ C\bigg{)}, \label{W} \f where $|I_L|$ is the amplitude of the
current through the inductor and $|U_C|$ is the amplitude of the voltage over the capacitor. Usually a patch
antenna is fed with a fixed voltage source, thus, $|U_C|$ is fixed. We see from (\ref{W}) that increase in $C$
leads to strongly increased stored energy, since we must at the same time decrease the value of $L$ to keep the
resonant frequency fixed. On the other hand, by increasing $L$ we can reduce $C$ and both factors decrease the
stored energy.

\subsection{The effect of static permeability and frequency dispersion}

Due to frequency dispersion of artificial magneto-dielectric substrates conventional \emph{dielectric substrates
are always better} in terms of retained impedance bandwidth in antenna miniaturization \cite{Pekka}. The effect
of frequency dispersion has a very fundamental nature, yet this effect is often neglected. A common misbelief is
that by selecting the operational frequency of the antenna closer to the substrate resonance the bandwidth is
more effectively retained since Re$\{\mu_{\rm eff}\}$ is larger. However, the closer one lies to the substrate
resonance, the steeper is the gradient of the dispersion curve.

To explain the effect of static permeability and frequency dispersion let us assume that the antenna shown in
Fig.~\ref{circuit}a.~is filled with a magneto-dielectric material obeying the modified Lorentzian type
dispersion in $\mu=\mu' - j\mu''=\mu'(1 - j\tan\delta_{\mu})$: \e \mu = \mu(\omega) = \mu_{\rm s} +
\frac{\Lambda\omega^2}{\omega_{\rm 0}^2 - \omega^2 + j\omega\Gamma}. \label{muLL} \f Above $\mu_{\rm s}$ is the
static permeability, $\Lambda$ is the amplitude factor ($0<\Lambda<1$), $\omega_{\rm 0}$ is the undamped angular
frequency of the zeroth pole pair (the resonant frequency of the medium), and $\Gamma$ is the loss factor. The
modified Lorentzian type dispersion rule is a commonly accepted model for dense arrays of split-ring resonators
and other similar structures (in this case $\mu_{\rm s}=1$), e.g.~\cite{Pendry, Sauviac, Maslovski}. Certain
composites containing natural magnetic constituents, e.g.~composite substrates containing thin ferromagnetic
films \cite{Lagarkov_mod}, can be modeled using the general Lorentzian model. Here we study for the sake of
clarity the effect of static permeability using (\ref{muLL}) since the general Lorentzian model reduces to
(\ref{muLL}) at low frequencies. In (\ref{muLL}) natural magnetism is described by $\mu_{\rm s}$, however,
(\ref{muLL}) is not in a general case applicable with natural magnetics (as will be explained in the next
section). In the vicinity of the magnetic resonance the effective permittivity of a dense array of split ring
resonators is weakly dispersive, and can be assumed to be constant. With composites containing ferromagnetic
films the effective permittivity can also be considered dispersion-free over a wide frequency range. Thus, we
assume $C = C_0\E$, where $\E$ is the effective permittivity of the substrate.

Usually a magnetic (magneto-dielectric) substrate is utilized well below the substrate resonance
(e.g.~\cite{Mosallaei_disagree,Buell,Pekka}) and we assume that $\omega_{\rm x}\ll\omega_{\rm 0}$. For the sake
of simplicity we assume that losses in the magnetic material are negligible in the vicinity of $\omega_{\rm x}$
($\Gamma=0$)\footnote{We discuss later in more detail situations when the loss contribution cannot be
neglected.}, in addition to this the loss tangent for the permittivity is assumed to be very small. With the
aforementioned assumptions the inductance reads \e L = L_0\mu(\omega) \approx L_0\mu_{\rm s}(1 +
\frac{\Lambda}{\mu_{\rm s}\omega_{\rm 0}^2}\omega^2) = L_0\mu_{\rm s}(1 + \alpha\omega^2). \label{Ld} \f The
resonant condition becomes \e 1 - \omega^2CL_0\mu_{\rm s}(1 + \alpha\omega^2) = 0, \label{rescond} \f and we can
readily solve the squared angular resonant frequency \e \omega_{\rm x}^2 = \frac{1}{2\alpha}\bigg{(}\sqrt{1 +
\frac{4\alpha}{CL_0\mu_{\rm s}}} - 1\bigg{)}. \label{modF} \f Dispersion in the material slightly changes the
resonant frequency compared to the case when there is no dispersion. When the antenna operates well enough below
the material resonance the dispersion is weak and the angular resonant frequency can be expanded into Taylor
series: \e \omega_{\rm x}^2 \approx \frac{1}{CL_0\mu_{\rm s}}\bigg{(}1 - \frac{\alpha}{CL_0\mu_{\rm s}}\bigg{)}.
\label{resf} \f With the assumptions described earlier the susceptance of the resonator reads (in the vicinity
of the antenna resonance) \e B = \omega{C} - \frac{1}{\omega{L_0}\mu_{\rm s}(1 + \alpha\omega^2)} \approx
\omega{C} - \frac{1}{\omega{L_0}\mu_{\rm s}}(1 - \alpha\omega^2). \label{suscp} \f The quality factor of the
resonator becomes \e Q = \frac{\omega}{2G}\frac{\partial{B}}{\partial{\omega}}\bigg{|}_{\omega = \omega_{\rm x}}
 = \frac{1}{2G}\bigg{(}\frac{1 + \omega^2CL_0\mu_{\rm s}}{\omega{L_0}\mu_{\rm s}} +
 \frac{\omega\alpha}{L_0\mu_{\rm s}}\bigg{)}\bigg{|}_{\omega=\omega_{\rm x}},
\label{Qgen} \f which can be further polished: \e Q = \frac{1}{G}\sqrt{\frac{C_0\E}{L_0\mu_{\rm s}}} +
\frac{\omega_{\rm x}}{2G}\frac{\alpha}{L_0\mu_{\rm s}} = Q^{\alpha=0} + \frac{\omega_{\rm
x}}{2G}\frac{\alpha}{L_0\mu_{\rm s}}. \label{Qder} \f Important observations can be made from (\ref{Qder}):
first of all, if there is no dispersion ($\alpha=0$) the quality factor reduces to the well known static result
given by (\ref{Qcircuit}). From causality we know that every passive and linear material is inevitably
dispersive (and lossy), thus in reality $\alpha\neq0$. Eq.~(\ref{Qder}) shows the strong effect of the static
permeability which was noticed already in \cite{Pekka}: if $\mu_{\rm s}=1$ the magnetic material only increases
the quality factor, no matter if the value of Re$\{\mu\}>1$ at the operational frequency of the antenna.
$\mu_{\rm s}>1$ is necessary to compensate the negative effect of frequency dispersion.

\subsection{Relative radiation quality factor}

The above derivation based on the lumped element model can be generalized to some extend by considering a
resonator whose spectrum is concentrated near a certain frequency $\omega_{\rm x}$ (the discussion partly
follows the discussion presented in \cite{Pekka}). Again we assume that the resonator is filled with
\emph{low-loss} materials $\E(\omega), \mu(\omega)$. Note that here we consider a general case of frequency
dispersion, the low-loss assumption is the only assumption that we make concerning the dispersion law in the
material. The known expression for the time-averaged energy density of electromagnetic fields reads
\cite{Landau,Jackson} \e w_{\rm em} = \E_0{
\partial[\omega\E(\omega)] \over
\partial\omega}\bigg{|}_{\omega=\omega_{\rm x}}{|E|^2\over 4} + \M_0
{\partial[\omega\M(\omega)] \over
\partial\omega}\bigg{|}_{\omega=\omega_{\rm x}}{|H|^2\over 4}.
\label{density} \f Above $|E|$ and $|H|$ are the amplitudes of the electric and magnetic fields. The energy
stored in the resonator can be found by integrating $w_{\rm em}$ over the volume of the resonator. From
(\ref{density}) we see that increase in both $\E$ and $\mu$ will increase the stored energy. In addition to
this, the increase of stored energy due to frequency dispersion is also explicit in (\ref{density}).

Let us consider a resonant $\lambda/2$-patch antenna having length $l$, width $w$, and height $h$. The antenna
element lies on top of a large, non-resonant ground plane. The radiation quality factor was derived in
\cite{Pekka} \e Q_{\rm r} = \frac{\pi{Y_0}}{8G_{\rm
r}}\bigg{[}\frac{1}{\mu(\omega)}\frac{\partial[\omega\mu(\omega)]}{\partial\omega}\bigg{|}_{\omega=\omega_{\rm
x}} + \frac{1}{\E(\omega)}\frac{\partial[{\omega\E(\omega)}]}{\partial{\omega}}\bigg{|}_{\omega=\omega_{\rm
x}}\bigg{]}. \label{Qdisp} \f Here $Y_0$ is the characteristic admittance of the patch segment
\cite{Wheeler_adm}, and $G_{\rm r}$ is the radiation conductance. If the material parameters are dispersion-free
$Q_{\rm r}$ given by eq.~(\ref{Qdisp}) is the same as considered e.g.~in \cite{Hansen}. In practise a patch
antenna is commonly fed using a short and narrow feed line to ease the impedance matching. For the discussion on
additional stored energy due to the feed strip, please see \cite{Pekka}.

When studying antenna miniaturization using novel antenna substrates, conventional dielectric substrates are
often used as a reference substrate. Usually with high accuracy the permittivity of the substrate can be
regarded dispersion-free over a wide frequency range. Let us in the reference case fill the volume under the
antenna element with a dielectric substrate offering the same size reduction [$\E^{\rm ref} = \mu(\omega_{\rm
x})\E(\omega_{\rm x})$]. The ratio between the radiation quality factors $Q_{\rm r}^{\rm rel}$ becomes \e
\frac{Q_{\rm r}}{Q_{\rm r}^{\rm ref}} =
\frac{1}{2\mu(\omega)}\bigg{[}\frac{1}{\mu(\omega)}\frac{\partial[\omega\mu(\omega)]}{\partial\omega}\bigg{|}_{\omega
= \omega_{\rm x}}+ \frac{1}{\E(\omega)}\frac{\partial[\omega\E(\omega)]}{\partial\omega}\bigg{|}_{\omega =
\omega_{\rm x}}\bigg{]}. \label{ratio} \f

Eq.~(\ref{ratio}) is the key formula allowing a proper comparison between the impedance bandwidth
characteristics of rectangular\footnote{The aforementioned steps of derivation can be repeated for different
antenna geometries.} patch antennas loaded with conventional dielectric substrates and novel antenna substrates.
It can be used when the following criteria are met: \begin{enumerate}
\item Both substrates have small losses.
\item The reference substrate does not possess any magnetic response and the permittivity can be considered
dispersion-free.
\item The voltage and current distribution remain unchanged, i.e.~both substrates fill the volume under the
patch element uniformly. \end{enumerate} Formula for the energy density [eq.~(\ref{density})] holds strictly
only when the absorption due to losses can be neglected. If losses in the material can not be neglected near the
frequency of interest, it is not possible to express energy density in terms of material permittivity and
permeability functions \cite{Landau}, pp.~28--30. One has to have knowledge about the material microstructure,
and this leads to a modification of the energy density expression \cite{Sergei_energy, Pekka2}. With artificial
magnetics dispersion is normal and loss contribution small at frequencies $\omega\ll\omega_{\rm 0}$. Thus, in
this case it is justified to neglect the loss contribution. However, for many microwave materials containing
natural magnetic constituents the dispersion is anomalous also well below the material cutoff frequency. In this
case the second term in the right part of (\ref{Qder}) is negative, and dispersion decreases the quality factor.
However, losses and dispersion appear together according to the low-frequency Kramers-Kronig relation \e
\mu''(\omega) \cong -\frac{\pi\omega}{2}\bigg{(}\frac{\partial\mu'(\omega)}{\partial\omega}\bigg{)}. \label{KK}
\f Losses in the substrate degrade the radiation efficiency of the antenna and increase the radiation quality
factor: by definition the radiation efficiency is the ratio between unloaded quality factor $Q_0$\footnote{The
quality factor in (\ref{Qder}) is the unloaded quality factor since only radiation losses are assumed.} and
$Q_{\rm r}$, thus, the radiation quality factor is calculated as \e Q_{\rm r} = Q_0/\eta_{\rm r}. \label{Qr} \f
Using the cavity theory $Q_{\rm r}$ can also be estimated as \e 1/Q_{\rm r} = 1/Q_0 - \tan\delta_{\mu} -
\tan\delta_{\E}, \label{Qr_mod} \f where $\tan\delta_{\mu}$ and $\tan\delta_{\E}$ are the magnetic and
dielectric loss tangents, respectively. As was shown in \cite{Sergei_Q} the use of anomalous dispersion is not
promising due to associated additional losses. In a practical situation the effect of anomalous dispersion is
negligible compared to the loss effect, and therefore the effect of anomalous dispersion is not considered in
the analysis of next section.

\section{Prospective magnetic materials for microwave frequencies: revision of the key properties}

Microwave performance of most magnetic materials containing natural magnetic constituents is due to the
ferromagnetic resonance. When modeling these materials the general Lorentzian type dispersion law can be used
for the complex permeability: \e \mu(\omega) = 1 + \frac{(\mu_{\rm s} - 1)\omega_{\rm res}^2}{\omega_{\rm res}^2
- \omega^2 + j\beta\omega\omega_{\rm res}}, \label{muR} \f where $\mu_{\rm s}$ is the static permeability,
$\omega_{\rm res}$ is the ferromagnetic (angular) resonant frequency, and $\beta$ is an empirical damping
factor. The complex permeability is close to the static permeability $\mu_{\rm s}$ at frequencies up to the
ferromagnetic resonant frequency $\omega_{\rm res}$, which can be considered as the permeability cutoff
frequency. The microwave permeability is large when both $\mu_{\rm s}$ and $\omega_{\rm res}$ are high. However,
it is well known that these values are related tightly to each other: higher $\mu_{\rm s}$ leads to lower
$\omega_{\rm res}$, and vice versa. For most bulk magnetic materials the relation is established by Snoek's law:
\e (\mu_{\rm s} - 1)\cdot{\omega_{\rm res}} = (4\pi/3)\gamma{4}\pi{M_{\rm s}}, \label{Snoek} \f where
$4\pi{M_{\rm s}}$ is the saturation magnetization and $\gamma\approx{3}$ GHz/kOe. In magnetic materials, both
$\mu_{\rm s}$ and $\omega_{\rm res}$ are dependent on the effective anisotropy field of the material, which, in
turn, depends on the composition, as well as on the manufacturing process, treatment, small admixtures, etc. The
right part of (\ref{Snoek}) is a function of the saturation magnetization, the value of which is determined by
the composition only. For this reason, Snoek's law is conventionally used for estimating the microwave
performance of the material. Within the limits imposed by Snoek's law, the static permeability and resonant
frequency can frequently be changed within a wide range by certain technological means.

In some occasions, instead of (\ref{Snoek}), the relation between the static permeability and resonant frequency
has the form \e (\mu_{\rm s} - 1)\cdot{\omega_{\rm res}}^2 = c(\gamma{4}\pi{M_{\rm s}})^2. \label{Snoekmod} \f
Eq.~(\ref{Snoekmod}) is valid for thin ferromagnetic films with in-plane magnetic anisotropy \cite{Acher}, and
in this case $c\approx4\pi^2$ and $\mu_{\rm s}$ is the largest component of permeability. The other case is
related to hexagonal ferrites \cite{Adenot} with $c=4\pi^2/3[1+(H_\theta + H_\phi)/(4\pi{Ms})]$, $H_\theta$ and
$H_\phi$ are out-of-plane and in-plane anisotropy fields respectively, and the material is assumed to be
isotropic. Only these two classes of magnets are known to obey (\ref{Snoekmod}). Eq.~(\ref{Snoekmod}) permits
higher static permeability compared to (\ref{Snoek}) with the same resonant frequency, provided that
$\omega_{\rm res}$ is not too high. The value in the right part of eq.~(\ref{Snoekmod}) is conventionally
referred to as Acher's constant of the material, $A$. However, this constant is typically derived from the
linear frequency rather than from the angular frequency. Therefore the right part of (\ref{Snoekmod}) is equal
to $4\pi^2A$, where $A$ is expressed in GHz$^2$.

In hexagonal ferrites, typical values of Acher's constant are not higher than 100 to 200 GHz$^2$
\cite{Adenot,Konst}. This means that with the resonant frequency $\omega_{\rm res}=2\pi\cdot{3}$ GHz the static
permeability can be as high as 10 to 20. For a discussion of hexagonal ferrites for miniaturization of
microstrip antennas, see \cite{Buell}. Ferromagnetic films with in-plane anisotropy may have much larger Acher's
constants due to higher saturation magnetization. For example, for iron $4\pi{Ms}=21.5$ kG and $A=3600$ GHz$^2$.
However, with iron films having thickness larger than approximately half a micrometer the microwave performance
is deteriorated by the effect of eddy currents. In addition to this, the out-of-plane component of magnetization
appears in thick films and also makes the microwave magnetic performance of the film worse.

To apply ferromagnetic films as a bulk material, which is needed e.g.~if the films are utilized as microstrip
antenna substrates, film laminates may be used \cite{Lagarkov_mod}. The laminates can be made using sputtering
process with a thin polymer substrate followed by stacking the sputtered substrates together to obtain a bulk
laminate. It is shown that for bulk laminates $c=4\pi^2p$ ($p$ is the volume fraction of magnetic components in
the laminate) in (\ref{Snoekmod}) \cite{Lagarkov1}. For sputtered microwave laminates made of iron, the film
thickness must not be larger than 0.3 mcm, and the thickness of substrate may be as low as 10 mcm
\cite{Lagarkov1}. Then $p = 3$ \% vol.~and $A=120$ GHz$^2$, a value which is about the same as that with
hexagonal ferrites. If several magnetic layers separated by non-magnetic interlayers are sputtered onto a
polymer substrate, the volume fraction $p$ may be increased by a factor of order of the number of layers
\cite{Lagarkov1}. With this method the potentially obtained microwave permeability values are the highest
possible among all magnets.

In some respect film laminates have advantages over ferrites. Ferrites are ceramic materials, hence they are not
easily machined (which can be necessary for accurate adjusting of the resonant size). Laminates consist mainly
of polymeric substrate, thus, they are easily machineable and may be shaped with high precision. Ferrites have a
specific weight of approximately 5 g/cm$^3$, while polymer-based film laminates weight about 1 g/cm$^3$. For
low-frequency antennas whose volume may be dozens of cubic centimeters, this difference may be essential.
Moreover, hexagonal ferrites have inherently high permittivity ($\E=15$ and larger \cite{Konst}), a feature
which is not typically desirable in antenna miniaturization in the view of retained impedance bandwidth. For
film laminates the permittivity is anisotropic: the component perpendicular to film layers, which is the
dominant component e.g.~if the laminates are used in microstrip antenna miniaturization, is as low as
approximately $\E=2...2.5$. Ferrites can be used as composites with ferrite powder to diminish the above
drawbacks, but in this case $A\propto{p}$ and $\mu_{\rm s}$ decreases correspondingly \cite{Konst}. On the other
hand, the film laminates require more complex technology and can therefore be more expensive. Another thing
which has to be remembered when using thin ferromagnetic films in antenna miniaturization is the conductivity
along the film layers. The effect occurs due to fringing electric fields and is seen as an increased effective
dielectric loss tangent. However, with patch antennas the effect can be reduced by leaving a short section of
empty space (length of the empty space about the height of the patch) near the radiating edges, and/or using
patterned films.

The most important issue deteriorating the microwave performance of the discussed materials is the effect of the
quality factor of ferromagnetic resonance. Most practically realizable materials have a low quality factor for
the ferromagnetic resonance, so that $\beta>1$ in eq.~(\ref{Snoek}) and $\mu'$ starts to decrease at frequencies
already well below the resonance. When $\beta\gg1$, the absorption peak and, correspondingly, the region of
stronger frequency dispersion are located in the vicinity of  $\omega_{\rm res}/\beta$ rather than in the
vicinity of $\omega_{\rm res}$. In this case, the permeability cutoff frequency is governed by Snoek's law
(\ref{Snoek}) rather than by Acher's law (\ref{Snoekmod}) \cite{Lagarkov_book}.

\begin{figure}[t!]
\centering \epsfig{file=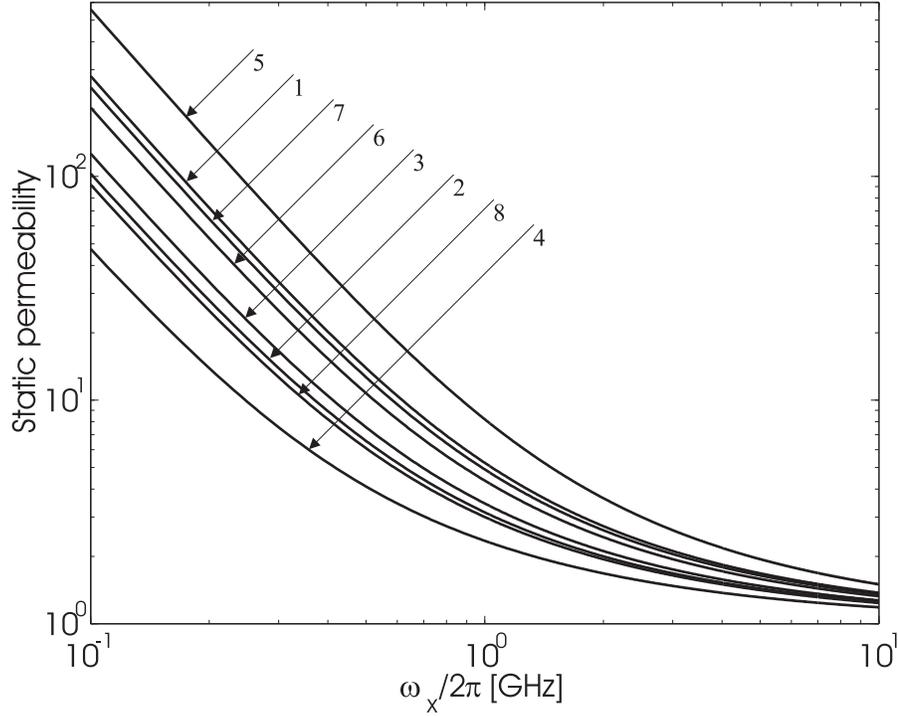, width=12cm} \caption{Obtainable static permeability as a function of
$\omega_{\rm x}$.} \label{ex}
\end{figure}

When the proposed materials are used in antenna miniaturization the most important effect of magnetic loss is
the rapid decrease of the radiation efficiency with increased frequency. From the low-frequency asymptote of the
Lorentzian dispersion law we arrive to the following bounding relation between the operating frequency of the
loaded antenna $\omega_{\rm x}$, the ferromagnetic resonance $\omega_{\rm res}$, the damping factor $\beta$, the
radiation quality factor $Q_{\rm r}$, and the radiation efficiency $\eta_{\rm r}$: \e \tan\delta_{\mu} =
\bigg{(}1 - \frac{1}{\mu_{\rm s}}\bigg{)}\frac{\beta{\omega_{\rm x}}}{\omega_{\rm res}}<\frac{1}{Q_{\rm
r}}\frac{1 - \eta_{\rm r}}{\eta_{\rm r}}. \label{restr} \f Above, the magnetic loss tangent is estimated based
on the Lorentzian dispersion law, in actual magnets additional loss mechanisms appear for low frequency loss
which may increase noticeably the loss tangent at very low frequencies compared to the resonance
\cite{Schloemann}. From the literature, fundamental restrictions concerning the quality factor of the
ferromagnetic resonance are not available. In actual samples, the quality factor is dependent on the
inhomogeneity of the sample, magnetic structure, etc. Most of the reported data on demagnetized polycrystalline
samples reveal wide-spread resonance curves. However, values of $\beta$ that are as low as 0.2 can be found for
both ferromagnetic films \cite{Acher2,Yamaguchi} and hexagonal ferrites \cite{Buell} (for all three data the
resonant frequency is 1.6 GHz). Generally, if $\omega_{\rm x}$, $Q_{\rm r}$, $\beta$, and the Acher's constant
$A$ are given, the ferromagnetic resonant frequency must be larger than \e f_{\rm res} > \frac{-2\cdot{3}^{1/3}A
+ 2^{1/3}(9AP + \sqrt{3}A\sqrt{4A + 27P^2})^{2/3}}{6^{2/3}(9AP + \sqrt{3}A\sqrt{4A + 27P^2})^{1/3}}, \label{gen}
\f with $P = (\omega_{\rm x}/2\pi)Q_{\rm r}\beta\eta_{\rm r}/(1 - \eta_{\rm r})$. Eq.~(\ref{gen}) is derived by
first solving $\mu_{\rm s}$ from (\ref{Snoekmod}) and substituting this into (\ref{restr}). By solving
$\omega_{\rm res}$ from (\ref{Snoekmod}) and substituting this into (\ref{restr}) one can study what kind of
static permeability values are possible given that $A,\beta,\omega_{\rm x},$ and $Q_{\rm r}$ are known. The
general expression is somewhat cumbersome, but in Fig.~\ref{ex} we present values for the static permeability as
functions of the operating frequency of the antenna $\omega_{\rm x}$. $A,\beta,$ and $Q_{\rm r}$ vary according
to Table \ref{tex}, $\eta_{\rm r}=50\%$ in every case. It is seen from Fig.~\ref{ex} that the obtainable static
permeability values decrease dramatically as the operating frequency of the antenna increases. If the antenna
operates in the GHz range the obtainable permeability values are rather moderate.

\begin{table}[b!]
\centering \caption{$A,\beta,Q_{\rm r}$ for different curves presented in Fig.~\ref{ex}.} \label{tex}
\begin{tabular}{|c|c|c|c||c|c|c|c|}
\hline
Curve no. & $A$ & $\beta$ & $Q_{\rm r}$ & Curve no. & $A$ & $\beta$ & $Q_{\rm r}$ \\
\hline
1 & 100 & 0.3 & 20 & 2 & 100 & 0.5 & 20 \\
3 & 100 & 0.3 & 30 & 4 & 100 & 0.5 & 30 \\
5 & 200 & 0.3 & 20 & 6 & 200 & 0.5 & 20 \\
7 & 200 & 0.3 & 30 & 8 & 200 & 0.5 & 30 \\
\hline
\end{tabular}
\end{table}

\section{Calculated impedance bandwidth properties with the proposed substrate}

\begin{figure}[b!]
\centering \epsfig{file=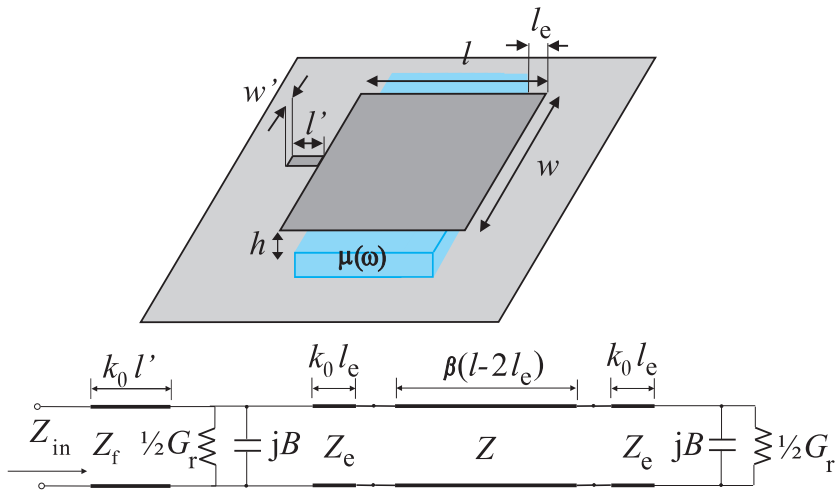, width=11cm} \caption{A schematic illustration of the antenna geometry and the
equivalent TL-representation of a strip fed patch antenna.} \label{sche}
\end{figure}

\begin{figure}[t!]
\centering \epsfig{file=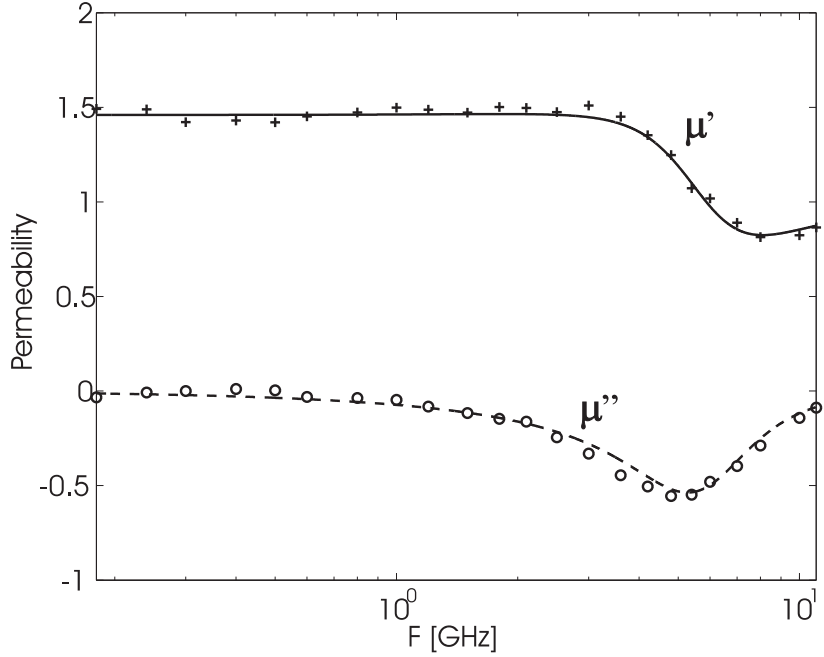, width=11cm} \caption{Dispersive behavior of $\mu$ of the proposed substrate.
Crosses and circles denote measured results. The sample has been produced and measured as described in
\cite{Lagarkov2}. The lines fit the experimental results to the Lorentzian dispersion model (\ref{muR}), and
give the following parameters: $\mu_{\rm s}=1.48$, $\omega_{\rm res}=2\pi\cdot{5.87}$ GHz, $\beta=0.903$.}
\label{mu}
\end{figure}

In the present section we model a strip fed $\lambda/2$-patch antenna lying on top of a non-resonant ground
plane using a TL-representation. A schematic illustration of the antenna and the equivalent TL-representation
are shown in Fig.~\ref{sche}. The formulation of the model is introduced in detail in \cite{Pekka}. The above
described layered substrate containing thin ferromagnetic films is chosen as the studied antenna substrate. The
dispersion of $\mu_{\rm eff}$ of the substrate is given in Fig.~\ref{mu}. The permittivity component
perpendicular to film layers is assumed to be dispersion-free and is estimated $\E_{\rm eff}=~2.0(1 - j0.01)$.
Moreover, we load the antenna volume with a reference dielectric substrate offering the same size reduction.

We have chosen 580 MHz for the center frequency of the miniaturization scheme. This frequency has strongly
increasing practical importance since it corresponds to the center frequency of the Digital Video Broadcast
(DVB) system \cite{DVB1}. As discussed in the previous section, for the practical utilization of composite
substrates containing natural magnetic inclusions the frequency should not be too high due to rapidly increasing
losses. For comparison, a demonstration for the applicability of a substrate containing sheets of Z-type
hexaferrite was conducted at 277 MHz \cite{Mosallaei_disagree}.

\subsection{Lossless substrate}

We start the analysis by calculating the impedance bandwidth properties when the antenna is loaded with lossless
substrates. We see that at 580 MHz $\mu_{\rm eff}'\cong1.461$. The value for the relative permittivity of the
reference substrate is found to be $\E_{\rm ref}'= 2.93$. The dimensions of the loaded patches are the
following: $l=w=163$ mm, $h=15$ mm, $l'=15$ mm. Feed strip width $w'$ is used to tune the quality of coupling
$T$ \cite{Pues} in different loading scenarios. We use a $-6$ dB matching criterion to define the impedance
bandwidth. The corresponding voltage wave standing ratio $S=3$. To allow a proper comparison between the
impedance bandwidth characteristics, $T$ has been tuned to $T=T_{\rm opt}=1/2(S + 1/S)$ by varying $w'$.

\begin{figure}[b!]
\centering \epsfig{file=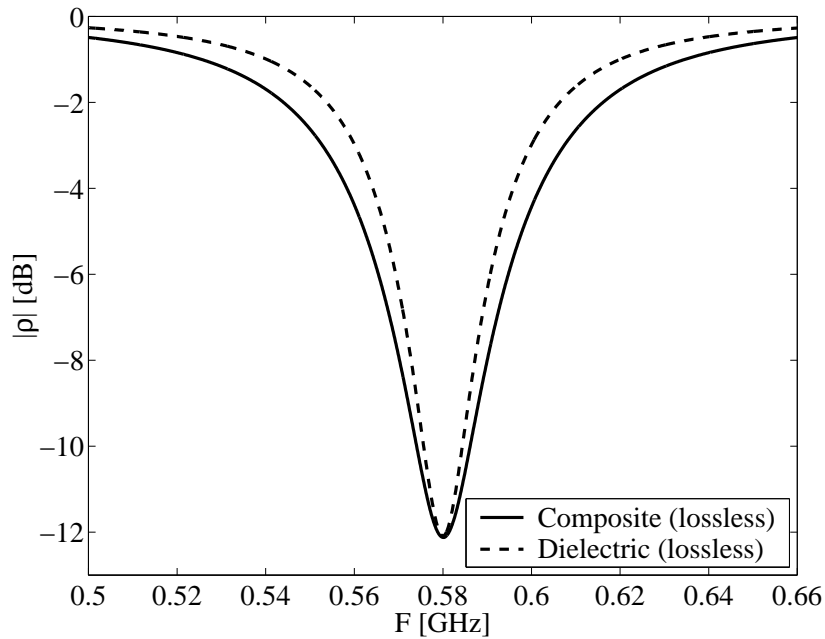, width=11cm} \caption{Calculated reflection coefficient with
different lossless substrates.} \label{S11_real_lossless}
\end{figure}

\begin{table}[t!]
\centering \caption{Calculated impedance bandwidth characteristics with lossless substrates.}
\label{Real_t_lossless}
\begin{tabular}{|c|c|c|c|c|}
\hline Loading & $BW\bigg{|}_{\rm -6 dB} [\%]$ & $Q_0$ & $\eta_{\rm r} [\%]$ & $Q_{\rm r}$ \\
\hline
Composite & 5.06 & 26.5 & 100 & 26.5 \\
Dielectric & 3.68 & 36.4 & 100 & 36.4 \\
\hline
\end{tabular}
\end{table}

Fig.~\ref{S11_real_lossless} shows the calculated reflection coefficients. The main calculated parameters are
gathered in Table \ref{Real_t_lossless}. The results presented in Fig.~\ref{S11_real_lossless} and Table
\ref{Real_t_lossless} agree well with the results given by the analytical expression for the relative radiation
quality factor: For the substrate used in the TL-model eq.~(\ref{ratio}) gives $Q_{\rm r}^{\rm rel}=0.69$, while
the TL-model predicts $Q_{\rm r}^{\rm rel}=0.73$. It is physically clear that the TL-model predicts a larger
$Q_{\rm r}^{\rm rel}$: The analytical expression does not take into account the feed strip which can be regarded
as an effective inductor \cite{Pekka}. Moreover, due to the feed strip usually a patch antenna operates slightly
above it's parallel resonance. This means that the standing wave pattern in the TL-model is not the ``optimal''
(parallel resonance) pattern for magnetic loading \cite{Pekka}.

The present result validates our previous discussion: The dispersion characteristics of the proposed substrate
are advantageous in antenna miniaturization. However, to draw a conclusion on the practical applicability of the
proposed substrate, realistic losses have to be taken into account.

\subsection{Practically realizable substrate}

\begin{figure}[b!]
\centering \epsfig{file=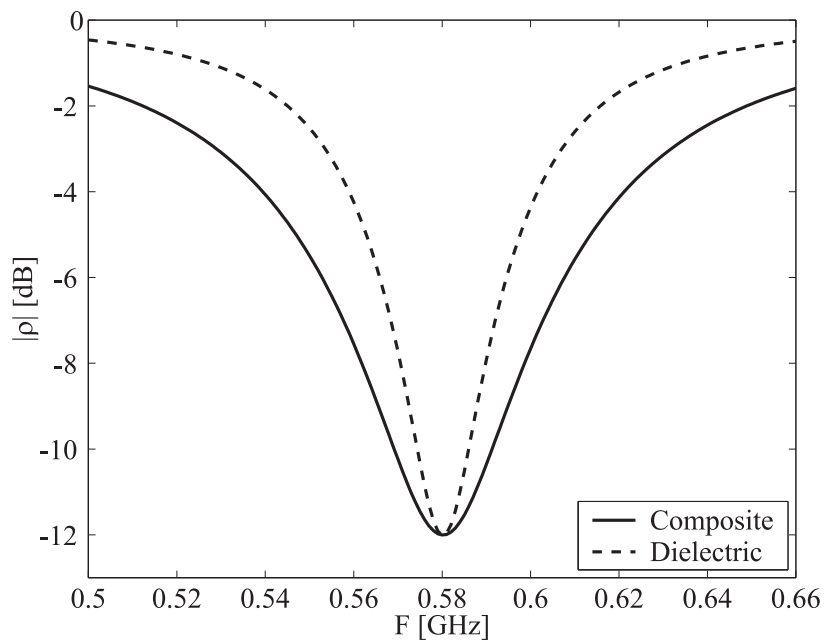, width=11cm} \caption{Calculated reflection coefficient with different
substrates. Practically realizable example.} \label{S11_real}
\end{figure}

The impedance bandwidth properties are calculated when the antenna is loaded with a practically realizable
composite substrate, Fig.~\ref{mu} [at 580 MHz $\mu_{\rm eff}\cong1.461(1 - j0.0284)$]. The relative
permittivity of the reference substrate is found to be $\E_{\rm ref} = 2.85(1 - j0.01)$. The dimensions of the
loaded patches are the following: $l=w=166$ mm, $h=15$ mm, $l'=15$ mm. Fig.~\ref{S11_real} shows the calculated
reflection coefficients. The main calculated parameters are gathered in Table \ref{Real_t}. $Q_{\rm r}^{\rm TL}$
denotes the radiation quality factor obtained using (\ref{Qr}), and $Q_{\rm r}^{\rm cav}$ denotes the radiation
quality factor obtained using (\ref{Qr_mod}). For the calculation of $Q_{\rm r}^{\rm TL}$ we calculate the
radiation efficiency using the numerical Wheeler Cap method \e \eta_{\rm r} = 1 - R/R_0, \label{etar} \f where
$R$ is the input resistance at the operational frequency when the radiation conductance is set to zero, and
$R_0$ is the input resistance at the operational frequency when the radiation resistance is that given by the
TL-model.

Losses in the substrate rather strongly degrade the radiation efficiency. The magnetic loss tangent is
$\tan\delta_{\mu}=0.0284$ at 580 MHz. Compared to the loss tangents of conventional high quality dielectrics the
value seems high. However, for comparison, the magnetic loss tangent of a substrate containing sheets of Z-type
hexaferrite was reported to be $\tan\delta_{\rm m}=0.019$ at 277 MHz \cite{Mosallaei_disagree}. Moreover, 580
MHz is a fairly high frequency for the utilization of a substrate containing natural magnetic constituents.
Based on the above discussion we consider the radiation efficiency \emph{promisingly high}. The most important
result is that the radiation quality factor is noticeably smaller than with pure dielectrics even in the
presence of realistic losses. This means that the antenna miniaturized using the proposed substrate operates
closer to the fundamental small antenna limit, thus its size-bandwidth characteristics are better.

\begin{table}[t!]
\centering \caption{Calculated impedance bandwidth characteristics. Practically realizable example.}
\label{Real_t}
\begin{tabular}{|c|c|c|c|c|c|}
\hline Loading & $BW\bigg{|}_{\rm -6 dB} [\%]$ & $Q_0$ & $\eta_{\rm r} [\%]$ & $Q_{\rm r}^{\rm TL}$ & $Q_{\rm r}^{\rm cav}$ \\
\hline
Composite & 9.45 & 14.15 & 48.4 & 29.2 & 31.0 \\
Dielectric & 4.94 & 27.08 & 73.8 & 36.7 & 37.2 \\
\hline
\end{tabular}
\end{table}

\begin{figure}[t!]
\centering \epsfig{file=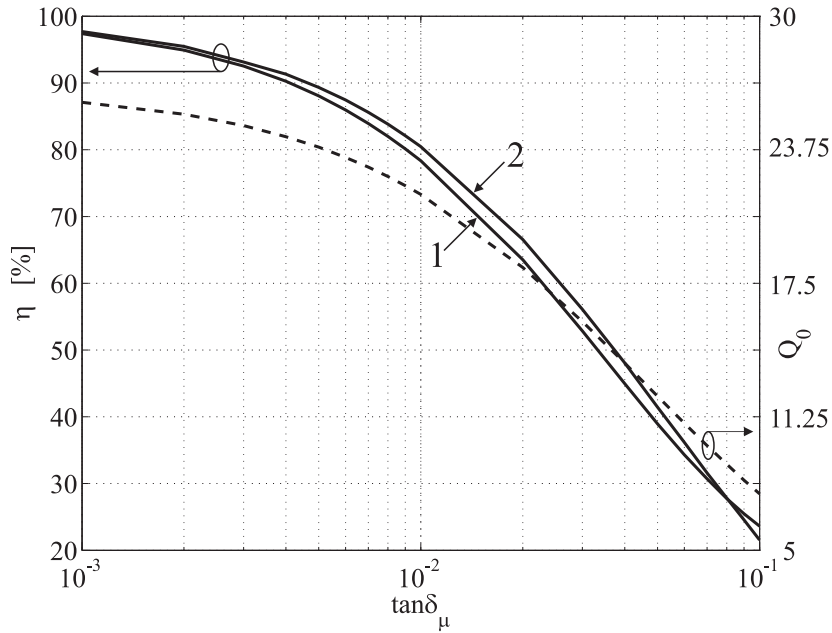, width=11cm} \caption{Calculated radiation efficiency (solid line)
and unloaded quality factor (dashed line) as a function of the magnetic loss tangent. Curve 1 corresponds to the
cavity theory estimation, curve 2 to the Wheeler Cap method. Dielectric loss tangent is assumed to be zero}
\label{efficiency}
\end{figure}
\subsection{Estimate for $\eta_{\rm r}$ as a function of magnetic loss tangent}

There is no theory for the magnetic loss tangent of composites filled with thin ferromagnetic films. Therefore,
we assume that with future technology the loss tangent can be brought down. Here we estimate what kind of
radiation efficiencies can be achieved with different magnetic loss tangents $\tan\delta_{\mu}$. We use two
different techniques to estimate the efficiency: 1) $Q_0$ is calculated using the TL-model and we use
(\ref{Qr_mod}) to estimate $Q_{\rm r}$ when the loss tangents are known. The radiation efficiency is then
calculated using (\ref{Qr}). 2) The numerical Wheeler Cap method is used to estimate the radiation efficiency.
When calculating $Q_0$ the dielectric loss tangent is assumed to be zero for the sake of clarity. With a given
$\tan\delta_{\epsilon}$ the radiation efficiency can conveniently be estimated using (\ref{Qr_mod}) and
(\ref{Qr}) when $Q_0$ and $\tan\delta_{\mu}$ are known. The result for the estimation is depicted in
Fig.~\ref{efficiency}.

\section{Conclusion}

In this paper we have discussed antenna miniaturization using magneto-dielectric substrates. The effect of
antenna substrate to the radiation quality factor has been presented. It has been shown that natural magnetic
constituents have to be embedded into the substrate in order to gain advantages over conventional dielectric
substrates. Detailed discussion has been presented concerning antenna substrates containing natural magnetic
inclusions. Moreover, we have introduced a layered substrate containing thin ferromagnetic films as a
prospective antenna substrate, and treated limitations for the obtainable permeability values. Using a
transmission-line model we have compared the impedance bandwidth properties of antennas loaded with the proposed
substrate and a conventional dielectric substrate. The results show that in terms of minimized radiation quality
factor the proposed substrate outperforms conventional dielectric substrates. Estimates for the radiation
efficiency are given as function of the magnetic loss tangent.

\section*{Acknowledgements}

We wish to thank Professor Constantin Simovski for useful discussions. K.~N.~Rozanov and A.~V.~Osipov are
grateful to Russian Foundation for Basic Research for partial support of the work according to grant
05--08--01212.

\end{document}